\documentclass[aps,prl,showpacs,floatfix,twocolumn,letterpaper]{revtex4}
\usepackage[pdftex]{graphicx}
\usepackage{amsmath}
\usepackage{hyperref}
\usepackage{latexsym}
\usepackage{color}
\usepackage{amssymb}
\usepackage{amsmath}

\begin{document}

\title{Controlling the hyperfine state of rovibronic ground-state polar molecules}

\author{S. Ospelkaus, K.-K. Ni, G. Qu{\'e}m{\'e}ner, B. Neyenhuis, D. Wang, M. H. G. de Miranda, J. L. Bohn, J. Ye, and D. S. Jin}
\affiliation{JILA, National Institute of Standards and
Technology and University of Colorado, and Department of
Physics,University of Colorado, Boulder, CO 80309-0440, USA}

\begin{abstract}
Ultracold molecules offer entirely new possibilities for the control of 
quantum processes  due to
their rich internal structure.  Recently, near quantum degenerate  gases of 
molecules have been prepared in their  rovibronic ground state.   For future 
experiments, it is crucial to also control their hyperfine state. 
Here, we report the preparation of a rovibronic ground state molecular quantum 
gas in a single hyperfine state and in particular in the absolute lowest 
quantum state. The demonstrated and presented scheme is  general for bialkali 
polar molecules and allows the preparation of molecules in a single hyperfine 
state or in an arbitrary coherent superposition of hyperfine states.  The 
scheme  relies on  electric-dipole, two-photon microwave transitions through 
rotationally excited states and makes use of electric nuclear quadrupole 
interactions to transfer molecular population between different hyperfine 
states. \end{abstract}


\maketitle

The field of ultracold atomic quantum gases draws much of its
success from the unprecedented ability to precisely control the
external and internal degrees of freedom of the gas. Control over
the external, or motional degree of freedom, comes from realizing
ultracold gases in almost arbitrary confining potentials provided by
magnetic or optical fields. The internal degrees of freedom, namely
the quantum states of the atoms, can be manipulated by driving rf~\cite{rf} or
optical transitions. Because collisional interactions
in the gas depend on the internal states, precise control of these
quantum states is a prerequisite for creating trapped samples that
are stable against inelastic collisions as well as for accessing
scattering resonances in order to tune the interparticle
interactions~\cite{Feshbach}. Manipulation of the internal degrees of freedom is
also essential in the study of quantum gases, where samples of
identical bosons or fermions in a single internal state can behave
very differently from spin mixtures. Finally, the precise control of
the atomic states is key to quantum information schemes where one
seeks to initialize and manipulate atoms as quantum qubits with long
coherence times~\cite{longcoh}.

The precise control of external and internal degrees of freedom will
be equally important for the emerging field of ultracold molecular
quantum gases. This field has recently seen
tremendous progress through the first preparation of near quantum
degenerate  gas of bialkali molecules in the rovibrational ground
state of the electronic ground molecular
potential~\cite{Ni08, Faraday}. These experiments have demonstrated a
high degree of control over the electronic, vibrational and
rotational degrees of freedom of ultracold molecules via two-photon
optical Raman transitions~\cite{STIRAP, STIRAPexp}.
However, most molecules will additionally
have hyperfine structure within a single rotational and vibrational
level~\cite{hyperfine}, and, as is true for ultracold atomic gases, control of these
quantum degrees of freedom is essential for future experiments. In particular, for experimental efforts to achieve a Bose-Einstein condensate (BEC) or quantum degenerate Fermi
gas of ground-state molecules it is advantageous to have all the
molecules in a single hyperfine state, and ideally in the hyperfine
state with the lowest energy.  For molecules, an additional
possibility is ultracold chemical reactions, whose study will
require reliable preparation and detection of the quantum state of
the molecules~\cite{Krems2008}.  While previous studies of chemical processes are
essentially all done in a temperature range where one has
statistical occupation of all hyperfine states, at the ultralow
temperatures recently achieved for ground-state polar
molecules the energy splitting between hyperfine
states can be much larger than typical collision energies in the
gas! Moreover, in the ultracold regime, quantum statistics play a significant role in collisional processes. Control over the nuclear spin of the molecules might even open the door to the study of SU(N) magnetism~\cite{sun2009}.

Here, we show that we can create ultracold KRb molecules
in a single hyperfine state.  In addition, we propose and
experimentally demonstrate a general scheme for controlling the
hyperfine state of bialkali polar molecules in their rovibrational
ground state.  This scheme allows for the production of ultracold
molecules in any arbitrary hyperfine state, and, in particular, is
used here to achieve an ultracold gas of KRb molecules all in their
absolute lowest energy state (electronic, vibrational, rotational,
and hyperfine).

\begin{figure}[tbp]
\begin{centering}
\leavevmode
\resizebox*{.85\columnwidth}{!}{\includegraphics{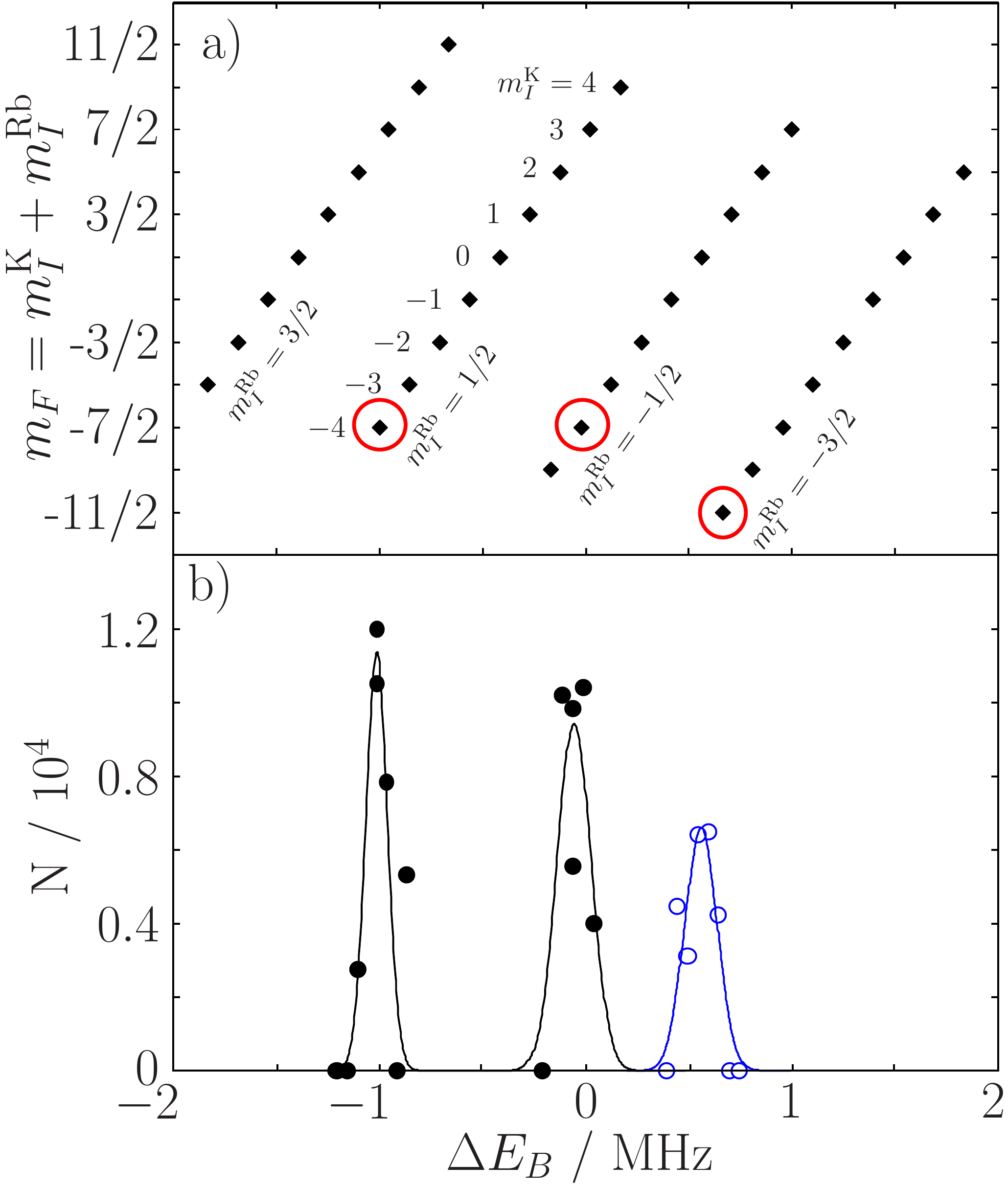}}
\end{centering}
\caption{({\bf color online}) (a) Hyperfine structure of
$^{40}$K$^{87}$Rb molecules in their rotational ($N=0$) and
vibrational ($v=0$) ground state of the electronic ground molecular
potential ($X^1\Sigma^+$) at  $B=545.9$\,G. The rovibrational ground
state splits into to 36 hyperfine states corresponding to the
different projections of the nuclear spins of the constituent atoms
$^{40}$K  ($I^K=4$) and $^{87}$Rb  ($I^{Rb}=3/2$). The circled quantum states can be directly populated
via STIRAP in our experiment. (b) Optical two-photon
spectroscopy of the rovibrational ground-state manifold of
$^{40}$K$^{87}$Rb molecules (see text). We observe three hyperfine
states, which are characterized by $m_F=-7/2$ (black points) and
$m_F=-11/2$ (blue open points). Here, $\Delta E_B$ is the binding energy of the hyperfine states at $B=546.9$\,G relative to the binding energy of the rovibrational ground state at zero magnetic field. Negative $\Delta E_B$ indicates a more deeply bound state and positive $\Delta E_B$ a less deeply bound state} \label{fig1}
\end{figure}

The lowest energy electronic potential for heteronuclear bialkali
molecules, such as KRb molecules, is a spin singlet state
(X${^1\Sigma^+}$) that has zero electronic angular momentum and spin.
Therefore, the hyperfine structure reduces to a nuclear substructure
due to the nuclear magnetic moments of the constituent atoms.
Figure~\ref{fig1}\,(a) shows a sketch of the hyperfine structure of
rovibrational ground state $^{40}$K$^{87}$Rb molecules at the
experimentally relevant magnetic field of  $B=545.9\,$G (see below).
The $^{40}$K atom has a nuclear spin of $I^K=4$, while $^{87}$Rb has
$I^{Rb}=3/2$; together, this gives a total of
$(2I^K+1)(2I^{Rb}+1)=36$ hyperfine states for the rovibrational
ground-state molecules.  At sufficiently large magnetic fields (typically $B>20\, \textrm{G}$),  the nuclear Zeeman effect
dominates over the coupling of the two nuclear spins to each other ($ c \vec {I^{\textrm{Rb}}}\cdot \vec {I^{\textrm{K}}}\ll \vec g_I^{\textrm{Rb,K}}\mu_B vec {I}^{\textrm{Rb,K}}\cdot \vec B$).
 We  therefore label the molecular hyperfine states with the  quantum numbers $m_I^\text{K}$ and $m_I^{\text{Rb}}$  describing the
projection of the nuclear spins onto the external magnetic-field
axis. \textit{A priori}, manipulation 
of nuclear spin states of the molecule is difficult because of the
weak coupling of the nuclear spin to external magnetic fields. In
addition, transitions between  quantum states with $\Delta
m_I=\pm 1$ are nearly degenerate, which
presents an obstacle to driving selectively a single nuclear transition.

We perform our experiments on a near quantum degenerate gas of polar
$^{40}$K$^{87}$Rb molecules in the rotational ($N=0$) and
vibrational ($v=0$) ground state of the electronic ground-state
molecular potential ($X^1\Sigma^+$)~\cite{Ni08}.   Starting from 
weakly bound $^{40}$K$^{87}$Rb Feshbach molecules formed in the
vicinity of a Fano-Feshbach resonance at $B=546.7\, $G~\cite{hetFesh, Zirbel2008}, we
subsequently transfer the molecules, at $B=545.9\,$G, into the
rovibrational ground state using a single step of  coherent
two-photon Raman transfer (STIRAP)~\cite{STIRAP}. The Raman process bridges an
energy gap of $h\cdot 125\,$THz.  The first question we consider is what
hyperfine state or states are populated in the Raman process? We
start with weakly bound Feshbach molecules prepared in a single
quantum state with a total angular momentum projection of
$m_F=-7/2$~\cite{Zirbel2008}. The two Raman laser beams are linearly polarized and
co-propagating parallel to the quantization axis defined by the
external magnetic field.  By conservation of angular momentum, the
resulting $X^{1}\Sigma^+\, (v=0, N=0)$ molecules are then restricted to
$m_F=m_I^{\text{K}}+m_I^{\text{Rb}}=-11/2, -7/2, -3/2$ ($\Delta m_F=\pm 2,0$).  Just this consideration leaves
8 possible hyperfine states for the ground-state molecules. However,
the hyperfine structure in the electronically excited state used in the Raman process will in general impose  additional selection rules. In the experiments, we have observed three accessible
hyperfine states. As seen in Fig. \ref{fig1}\,(b), we can
spectroscopically resolve the hyperfine states for a given $m_F$
value.  Based on two-photon selection rules and the binding energies of the hyperfine states, we can identify the states and assign the quantum numbers $m_F=-7/2$ and 
$m_F=-11/2$. In particular, the fact that the highest energy state in Fig.
\ref{fig1}\,(b) disappears for spectra taken with identical
circular polarization for both Raman laser beams allows us to
identify this state as $m_F=-11/2$. 

Thus, we conclude that our two-photon Raman process, starting from a
well-defined Feshbach molecule state, selectively populates a single
hyperfine state within the rovibrational ground state manifold
($X^1\Sigma^+, v=0, N=0$). However, this state is not the lowest energy
hyperfine state of the molecule.  In the remainder of the paper, we
address the question of how to manipulate the hyperfine state of our
ground-state polar molecules and, ideally, put them in any one of
the 36 possible hyperfine states.  In addressing this question, we
first prepare molecules in the lowest energy state in Fig. \ref{fig1}\,(b),
which is characterized by $m_I^\text{K}=-4$ and $m_I^\text{Rb}=1/2$,
as a starting point for further manipulation of the hyperfine state.

\begin{figure}[tbp]
\begin{centering}
\leavevmode
\resizebox*{1.0 \columnwidth}{!}{\includegraphics{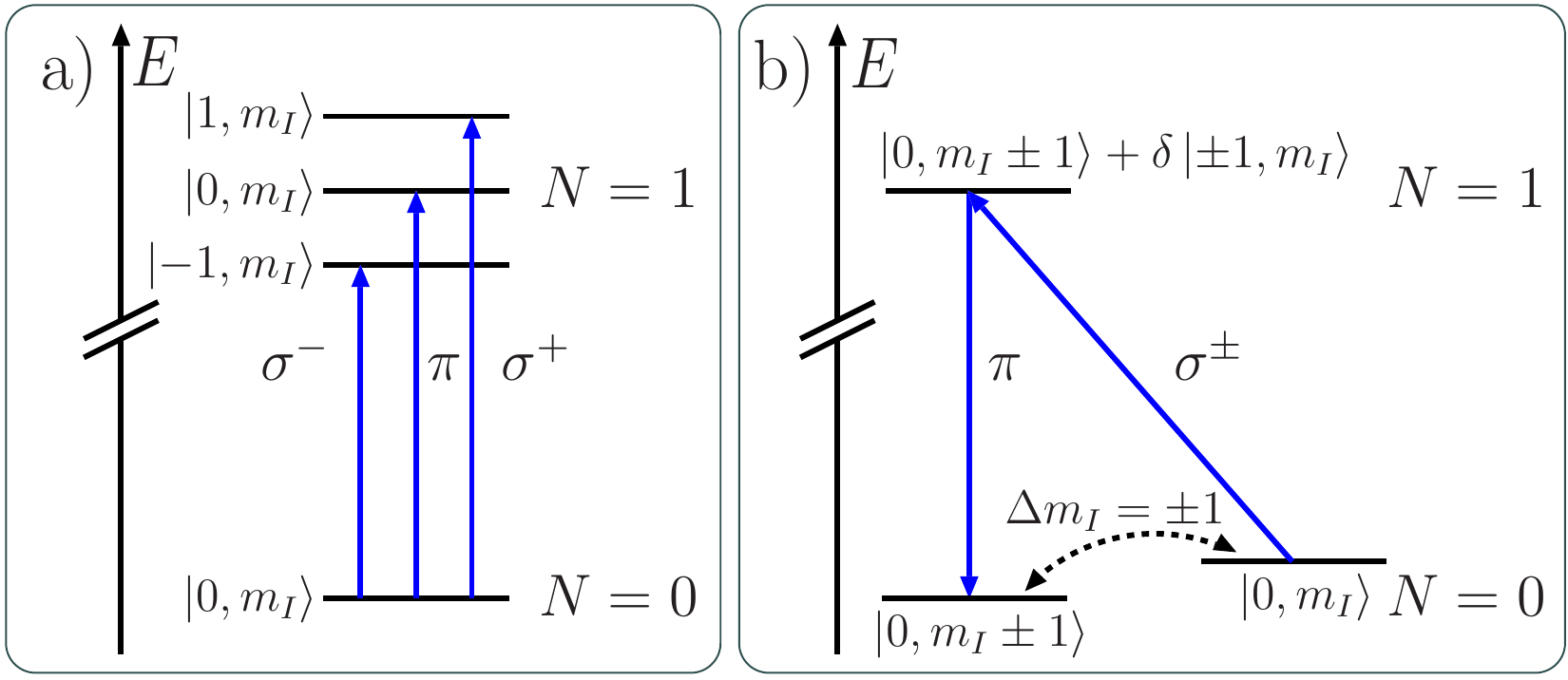}}
\end{centering}
\caption{({\bf color online}) (a) Electric dipole transitions
between the rotational ground state $N=0$ and the first
rotationally excited state $N=1$. To first order, electric dipole
transitions will keep the hyperfine state unchanged $\Delta m_I=0$.
(b) Controlling the hyperfine state within the rotational and
vibrational ground state. The interaction between the nuclear
electric-quadrupole moment of $^{40}$K / $^{87}$Rb and the electric
field gradient of the electronic cloud mixes quantum
states with different nuclear spin quantum number
$m^{\mathrm{K/Rb}}_I$ in $N=1$. This allows the implementation
of a two-photon scheme to transfer molecules between different
hyperfine states within the rovibrational ground state ($\Delta
m_I^{K/Rb}=\pm 1$). To simplify the notation, we abbreviate
$|N,m_N,m_I^\text{K}, m_I^{\text{Rb}}\rangle$ by $N$ $|m_N,
m_I\rangle$. Here $m_I$ refers to the changing hyperfine quantum
number $m_I$ of K or Rb.      } \label{fig2}
\end{figure}

We  manipulate the hyperfine state of the ground-state
molecules using microwave transitions that couple the rotational
ground state to a rotationally excited state.  Figure~\ref{fig2}
illustrates the basic idea for a two-photon microwave transfer
scheme within the hyperfine structure of the rovibrational ground
state.  The permanent electric dipole moment of 
polar molecules allows strong transitions  between the rotational
ground state and the first rotationally excited state.  The
single-photon electric dipole transition has a Rabi frequency given
by  $\frac{\vec d \cdot \vec {\mathcal E}}{\hbar}$, where $\vec d$ is
the permanent electric dipole moment of the molecules and $\vec{
\mathcal E}$ is the electric-field of the microwave radiation.  To
leading order, these electric dipole transitions are  only allowed  between different rotational states ($\Delta N=\pm 1$,
$\Delta m_N=\pm1,0$) while leaving the nuclear spins unchanged (see
Figure~\ref{fig2}\,(a)). Here, $N$
is the rotational quantum number and $m_N$ is the projection of the
rotation on the external magnetic-field axis. However, there is a higher order effect due to
the interaction between the nuclear electric-quadrupole moments of
$^{40}$K and $^{87}$Rb and the electric-field gradient created by
the electrons at the nuclear positions.  This interaction couples
the rotation and the nuclear spins, and allows one to change the
hyperfine state of the rovibrational ground-state molecule via two
microwave transitions (driving up to a rotationally excited state
and then back down to a different hyperfine ground state).

In the following, we will work with an uncoupled basis set of the
form $\left|N,  m_N,  m_I^\text{K},  m_I^\text{Rb} \right> $. For an $N=0$ state, the
electric-quadrupole interaction vanishes, and ignoring nuclear
spin-spin interactions, this basis set corresponds to the hyperfine
eigenstates.  For rotationally excited states such as $N=1$, the
electric-quadrupole interaction mixes quantum states of different hyperfine character with the same sum $m_N+m_I^\textrm{Rb}$, which results in eigenstates of the
molecular Hamiltonian of the form $ \left |N=1, m_N=0,
m_I^\text{K}, m_I^{\text{Rb}}\pm 1\right >+\delta \left
|N=1,m_N=\pm 1,m_I^{\text{K}}, m_I^{\text{Rb}} \right> $. (For
simplicity, we restrict the discussion here to states relevant for
manipulating  $m_I^{\text{Rb}}$,
however, $m_I^{\text{K}}$ can be manipulated in a similar manner.) 
 Here, the parameter $\delta$ typically is small ($|\delta |^2\ll 1$). 
Starting from a particular hyperfine state within the rovibrational
ground state,  $\left|0, 0, m_I^\text{K},  m_I^\text{Rb}\right >$,
  one can drive a microwave transition to a rotationally excited state
with predominantly different hyperfine character, such as $\left |1,
0, m_I^\text{K}, m_I^\text{Rb}\pm 1 \right > $, therefore
changing the nuclear spin of Rb by $\Delta m_I^{\text{Rb}}=\pm 1$.
  The Rabi frequency of the hyperfine
changing transition is then simply given by $|\delta|\frac{\vec d
\cdot \vec{\mathcal{E}}}{\hbar}$. Combining this hyperfine changing
microwave transition with a second microwave transition that
preserves the nuclear spin and transfers the molecules back into the
rotational ground state manifold ( $\left |1, 0, m_I^\text{K},
m_I^\text{Rb}\pm 1 \right > $ to $\left |0, 0, m_I^\text{K},
m_I^\text{Rb}\pm 1 \right > $), we can effectively transfer
molecular population within the rovibrational ground-state hyperfine
manifold (see Fig.~\ref{fig2} (b)). The selection rules for the
two-photon transfer scheme are then given by $\Delta m_I^{\text
K/Rb}=\pm 1$, and this process can be repeated as needed to prepare
the molecules in any hyperfine state within the rovibrational
ground-state manifold.

\begin{figure}[tbp]
\begin{centering}
\leavevmode
\resizebox*{0.85\columnwidth}{!}{\includegraphics{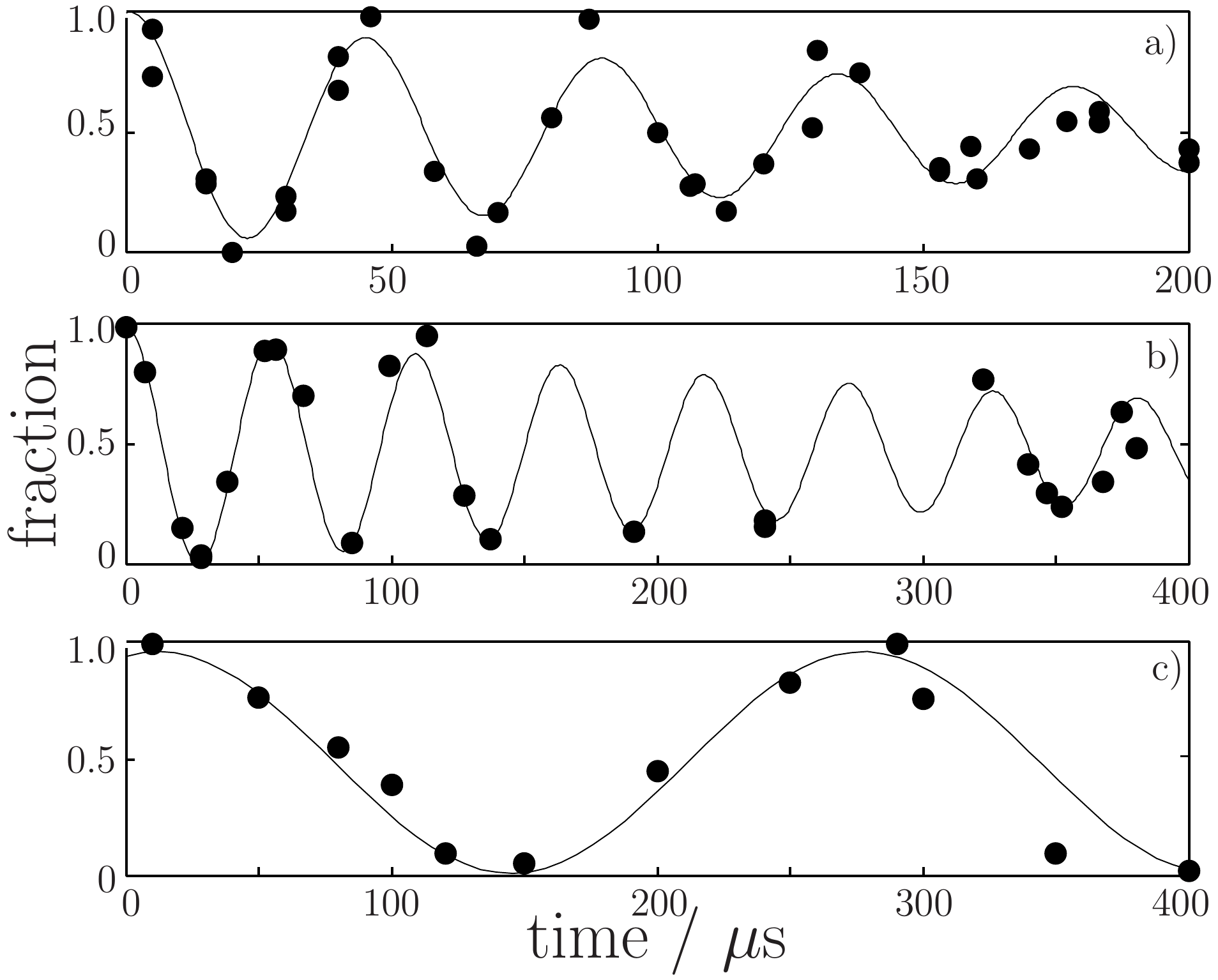}}
\end{centering}
\caption{ Rabi oscillations on a (a) hyperfine preserving microwave
transition $|0,0,-4,1/2\rangle \rightarrow |1,1,-4,1/2\rangle$  (b)
Rb hyperfine changing transition $|0,0,-4,1/2\rangle \rightarrow
|1,0,-4,3/2\rangle+\delta|1,1,-4,1/2\rangle$  (c) K hyperfine
changing transition $|0,0,-4,1/2\rangle \rightarrow
|1,0,-3,1/2\rangle+\delta|1,1,-4,1/2\rangle$. Note the different time axis in panel (a). The microwave power was reduced by a factor of 4 for the data in (a) resulting in an effective decrease of the Rabi frequency by a factor of 2.  } \label{fig3}
\end{figure}

\begin{table}[tbp]
\begin{center}
\begin{tabular}{|c|c|c|c|c}
\hline
 main state &  admixed state & $|\delta_\text{th}|^2$ &
$|\delta_\text{exp}|^2$   \\ [0.5ex]
\hline
 $|1 \ 0 \ \text{-}4 \ 3/2 \rangle$  & $|1 \ 1 \ \text{-}4 \ 1/2 \rangle$ &
0.045 & $\approx 0.1$  \\
\hline
 $|1 \ 0 \ \text{-}3 \ 1/2 \rangle$  & $|1 \ 1 \ \text{-}4 \ 1/2 \rangle$ &
0.0085 & $\approx 0.0064$  \\
\hline
\end{tabular}
\end{center}
\caption{Mixing of different hyperfine states in the rotationally
excited state $N=1$ at $B=545.9$~G. We compare the theoretically
calculated admixture $|\delta_\text{th}|^2$ to the experimentally
measured admixture $|\delta_\text{exp}|^2$. \label{TAB1} }
\end{table}

\begin{table}[tbp]
\begin{center}
\begin{tabular}{|c| c| c| c| c|}
\hline State $i$ & state $j$ & exp.   & theory   &
abs. diff.       \\ [0.5ex] \hline
\hline
$|0 \ 0 \ \text{-}4 \ 1/2 \rangle$ & $|1 \ 1 \ \text{-}4 \ 1/2 \rangle$ &
2227837(5)& 2227835 & 2  \\
$|0 \ 0 \ \text{-}4 \ 1/2 \rangle$ & $|1 \ 0 \ \text{-}4 \ 1/2 \rangle$ &
2228125(5)& 2228119& 6 \\
$|0 \ 0 \ \text{-}4 \ 1/2 \rangle$ & $|1 \ \text{-}1 \ \text{-}4 \ 1/2
\rangle$ & 2227774(7)& 2227776& 2  \\
$|0 \ 0 \ \text{-}4 \ 1/2 \rangle$ & $|1 \ 0 \ \text{-}4 \ 3/2 \rangle$ &
2227009(2)& $2227008$ & 1    \\
$|0 \ 0 \ \text{-}4 \ 1/2 \rangle$ & $|1 \ \text{-}1 \ \text{-}4 \ 3/2
\rangle$ & 2227133(20)& 2227128& 5  \\
$|0 \ 0 \ \text{-}4 \ 1/2 \rangle$ & $|1 \ 0 \ \text{-}3 \ 1/2 \rangle$  &
2228237(10)& 2228225& 12   \\
$|0 \ 0 \ \text{-}4 \ 1/2 \rangle$ & $|1 \ 1 \ \text{-}4 \ \text{-}1/2
\rangle$ & 2228588(5) &  2228593& 5  \\
$|0 \ 0 \ \text{-}4 \ 1/2 \rangle$ & $|1 \ 0 \ \text{-}4 \ \text{-}1/2
\rangle$ & 2228804(1)& 2228805& 1  \\ [1ex]
\hline
$|0 \ 0 \ \text{-}4 \ 3/2 \rangle$ & $|1 \ 0 \ \text{-}4 \ 3/2 \rangle$ &
2227765(10)& 2227761& 4\\ [1ex]
\hline
$|0 \ 0 \ \text{-}3 \ 1/2 \rangle$ & $|1 \ 0 \ \text{-}3 \ 3/2 \rangle$ &
2228109(16)& 2228091 & 18  \\ [1ex]
\hline
\end{tabular}
\end{center}
\caption{ Spectrum of rotational transitions from hyperfine state
$|i\rangle$  within $N=0$ to hyperfine state $|j\rangle$ within the
$N=1$ manifold.  We compare the experimentally measured transition
frequencies to the theoretical calculation. All frequencies are given in kHz.
\label{TAB2} }
\end{table}

The efficiency of the scheme relies critically on the strength of
the mixing of different hyperfine basis states in the rotationally
excited state $N=1$. To evaluate the strength of the mixing, we
diagonalize the total Hamiltonian ${\cal H}$ of a
$^{40}$K$^{87}$Rb$(v=0,N)$ molecule \cite{BrownCarrington03}
in the uncoupled basis $|N, \ m_N, \ m_I^\text{K}, \
m_I^\text{Rb} \rangle $, in the presence of an external magnetic
field of $B=545.9\,~$G,  and  calculate the 36 (108) eigenvalues and
eigenstates of the hyperfine structure within the
rotational $N=0$ ($N=1$) manifold.
We use the molecular parameters available in the literature~\cite{Ni08,Aldegunde08,Stone05}.

Experimentally, the mixing of different hyperfine basis states in the
rotationally excited state  $N=1$ can be measured by driving
hyperfine changing microwave transitions  from $N=0$ to $N=1$ and comparing
the strength of these transitions to the corresponding nuclear spin
preserving transitions. Figure~\ref{fig3} shows a comparison between
Rabi oscillations for three different microwave transitions: a nuclear spin
preserving transition ($\Delta m_I^{\textrm{K,Rb}}=0$, Fig.~\ref{fig3}\,(a) and  hyperfine changing transitions within the Rb ($ m_I^{\textrm{Rb}}$) and K  ($ m_I^{\textrm{K}}$) spin manifold, respectively (Fig.~\ref{fig3}\, (b) and (c)).

Table~\ref{TAB1} summarizes  the mixing parameter $|\delta|^2$ for
these particular hyperfine state combinations.
The mixing between different hyperfine states related to the K nuclear spin is typically $<1\%$, whereas it is almost  an order of magnitude larger for Rb hyperfine states. This reflects the large quadrupole moment of the Rb nucleus.  Note  that the above  investigated microwave transitions from $N=0$
to $N=1$   demonstrate spin flips within both the nuclear
structure of Rb and K ($\Delta m_I^\text{Rb/K}=\pm 1$), which allows  the preparation of arbitrary hyperfine states within the rovibrational ground state.
Using this scheme, we  also demonstrate the preparation of a
molecular cloud in the  lowest hyperfine state of the
rovibrational ground state manifold characterized by the quantum
numbers  $\left|N=0,m_N=0, m_I^\text{K}=-4,
m_I^{\text{Rb}}=3/2\right>$ (see Fig.~\ref{fig1} (a)).
An independent theoretical calculation of the microwave spectra of rovibrational ground-state polar $^{40}$K$^{87}$Rb molecules is available in~\cite{Aldegunde09}.

Finally, the experimentally measured one-photon microwave
spectrum starting from $|N=0, m_N=0, m_I^{\text{K}}=-4,
m_I^\text{Rb}=1/2\rangle$ to the rotationally excited manifold $N=1$
is fitted to  the theoretical calculation of the spectrum. We use the fit for a measurement of three  molecular parameters -
the rotational constant $B$, 
and the electric-quadrupole $eqQ_\textrm{K}$ and $eqQ_\textrm{Rb}$ 
which have previously only been predicted by \textit{ab initio} calculations~\cite{Aldegunde08}.
Table~\ref{TAB2} summarizes the results. 
From the best fit,
we determine that $B = 1.113950(5)$~GHz, $eQq_\textrm{K} = 0.45(6)$~MHz
and $eQq_\textrm{Rb} = -1.41(4)$~MHz. Our results are consistent with the \textit{ab initio} calculations in~\cite{Aldegunde08}, but  reduce significantly  the uncertainty on the molecular parameters as compared to~\cite{Aldegunde08} .

To summarize, we have demonstrated  a general scheme for controlling the
hyperfine state of bialkali polar molecules in their rovibronic
ground state.  This scheme allows for the production of ultracold
bialkali molecules in any arbitrary hyperfine state, and  in particular,  for the preparation of an ultracold gas of KRb molecules in their
absolute lowest energy state (electronic, vibrational, rotational,
and hyperfine).This paves the way for quantum-state controlled studies of elastic, inelastic, and chemically reactive collisions of polar molecules in the ultracold regime.

\begin{acknowledgments}
We acknowledge financial support from NIST, NSF and DOE. K.-K.N. and B.N. acknowledge support from NSF, S.O. from the Alexander-von-Humboldt Foundation, and M.H.G. de M. from CAPES/Fulbright. We thank J. Perreault, M. Swallows and C. Ospelkaus for critical reading of the manuscript.
\end{acknowledgments}

\end{document}